\documentclass[aps,prl,preprint,superscriptaddress]{revtex4}

\usepackage{graphicx}
\usepackage{dcolumn}
\usepackage{bm}

\begin{document}


\title{Quantum Key Distribution based on Single Photon Bi-partite Correlation}




\author{Kim Fook Lee}
\email[]{kflee@mtu.edu}
\author{Yong Meng Sua}
\affiliation{Department of Physics,\\ Michigan Technological
University,\\ Houghton, Michigan 49931}
\author{Harith B. Ahmad}
\affiliation{Department of Physics, \\ University of Malaya,\\ 50603 Kuala Lumpur, Malaysia}


\date{July 20, 2012}

\begin{abstract}
We present a scheme for key distribution based on bi-partite
correlation of single photons. Alice keeps an ancilla photon and
sends a signal photon to Bob, where intrinsic bi-partite correlation
of these photons is obtained through first order intensity
correlation in their detectors. The key bits are distributed through
sharing four bi-partite correlation functions and photon counting.
The scheme consists of two parts; first, Alice prepares
deterministic photon states and Bob measures the photon states based
on his random choice on correlation functions. Second, Alice guesses
Bob's choice of correlation functions and sets the key bits by
sending out photon states. Bob verifies the key bits through the
photon states regardless Alice made a right or wrong guess. We
called this key distribution scheme as prepare-measure-guess-verify
(PMGV) protocol. We discuss the protocol by using a highly
attenuated laser light, and then point out the advantages of using a
fiber based correlated photon-pair to achieve better performance in
security, communication distance and success rate of key
distribution.

\end{abstract}

\maketitle


Superposition and entanglement are essential in developing quantum
information technologies for real world applications especially for
secure communication with quantum key distribution
(QKD)~\cite{Gisin02,Scarani09}. QKD has been securely implemented in
an optical free-space link (144 km) with polarization entangled
photon-pair~\cite{zeilingerFS} and an optical fiber network (45 km)
with six different protocols~\cite{zeilingerOE}. The BB84 and B92
protocols~\cite{BB92a, BB92b} are secure against photon number
splitting attack (PNS). This attack is only a threat if Alice and
Bob shared a fake single-photon source such as using a highly
attenuated laser light. To overcome photon number splitting attack
(PNS)~\cite{Brassard00,Lutkenhaus00} due to the use of highly
attenuated laser light in a lossy long channel, decoy-state
protocols~\cite{Hwang03,Lo05,Wang05} and SARG04
protocol~\cite{Scarani04} have been proposed and implemented.
Measurement-device-independent QKD~\cite{LoPRL12} is recently
proposed to enhance secure communication against all detector side
channel attacks and double communication distance by using highly
attenuated laser light. The approach requires coincidence detection
of a signal pulse from Alice and a reference pulse from Bob. In this
work, we propose a QKD scheme based on single photon bi-partite
correlation.

Correlation functions or expectation values of two observable are
classical information as a consequence of the collapse of wave
functions in a measurement process. Correlation functions of two
Einstein-Podolsky-Rosen (EPR) entangled photons are obtained through
nonlocal interferences of their probability amplitudes at two
observers. The quantumness of these interferences is the exhibition
of particle-wave duality. The quantum mechanics without probability
amplitude was proposed~\cite{wootters86}, leading to the possibility
of quantum information processing without probability amplitudes,
that is, quantum information processing with correlation functions.

Bi-partite correlation of coherent light state has been observed by
wave mechanical interferences of electromagnetic light fields with
different modulated frequencies~\cite{kflee02,kflee04}. By
post-selecting a pair or multiple pairs of beat frequencies from
detectors, the bi-partite or multiple-partite correlation functions
were obtained for simulating the violation of Bell's
inequalities~\cite{aspect81} and the locality of
Greenberger-Horne-Zeilinger (GHZ) entanglement~\cite{JWPan00,Dik99}.
Recently, a coherent light field and a random phase-modulated noise
field was used to generate the optical bi-partite correlation
without applying any post-selecting techniques~\cite{kflee09}.
Instead, the correlation was obtained through mean-value measurement
of the multiplied random beat signals of two observers. The
experiment showed that the phases information between two observers
was not diminished in the presence of random noises. We further
interrogated the generation of bi-partite correlation by using two
weak coherent states in balanced homodyne detection~\cite{sua11},
where quantum noises, shot noises and electronic noises were
included in the measurement process. These noises were used to
protect the phases information between two observers. Bits
correlations were then extracted from the correlation functions.

In this paper, we present a new protocol for key distribution by
using single photon bi-partite correlation. The security of the
protocol is protected by the principle of quantum mechanics such as
quantum non-cloning theorem, superposition and entanglement. We use
four bi-partite correlations to distribute key between Alice and
Bob. Our protocol is relied on interference of single photons, i.e.,
the first order intensity correlation of an ancilla photon in Alice
and a signal photon in Bob. These single photons have to be
intrinsically correlated through a highly attenuated laser light or
a photon-pair source. We will first discuss the protocol by using
single photons (ancilla and signal) prepared from a highly
attenuated laser light. Our protocol requires coincidence detection
of ancilla photon and signal photon. We then discuss the use of a
fiber based correlated photon-pair or two time-synchronized
deterministic single-photon sources for improving the success rate
of key distribution and increasing the communication distance by
factor two in comparison to the use of a highly attenuated laser
light. The correlated photon-pair is easy-to-use and more tolerant
to decoherence compared to an entangled photon-pair. The essence of
the paper is to propose a new scheme for key distribution based on
single photon bi-partite correlation. Secure communication between
Alice and Bob is established through four types of bi-partite
correlation functions (C1, C2, C3, C4), which can be divided into
two groups $(\Psi, \Phi)$. Alice first prepares a sequence of
deterministic states, Bob measures each photon state randomly based
on his random choice on C1, C2, C3, or C4 and then tells Alice
through classical channel about the sequence of groups $(\Psi,
\Phi)$ that he has randomly chosen. Alice guesses the correlation
based on the group information and sets the key bit by sending
another signal photon to Bob. Bob verifies the key bit by the
outcome of his measurement regardless Alice made a right or wrong
guess. We called this scheme as prepare-measure-guess-verify (PMGV)
protocol.

\begin{figure}
\includegraphics[scale=0.7]{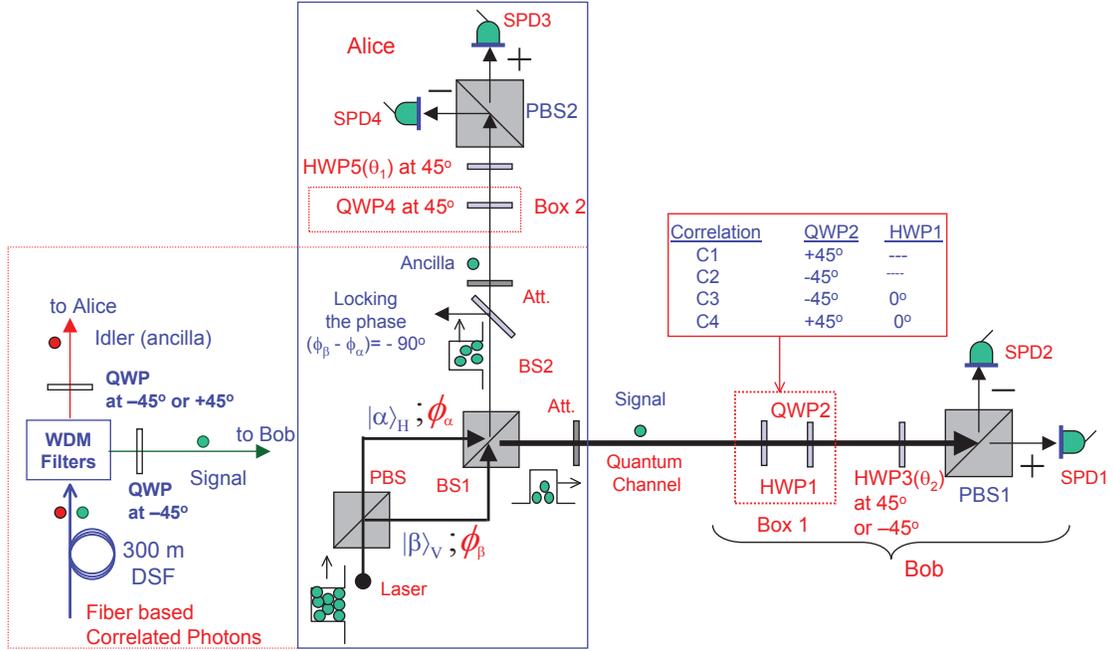}
\caption{The experiment scheme for implementing key distribution
using a signal photon and an ancilla photon. Also shown is the use
of a fiber based correlated photon-pair as photon source to replace
the highly attenuated laser light. The dotted box is the wave plates
used for preparing bi-partite correlation between Alice and Bob.}
\end{figure}

The proposed experiment setup is shown in Fig.1. A pulsed,
$45^{\circ}$-polarized laser light is used to provide a coherent
state with large mean photon number per pulse. The coherent state is
split by a polarizing beam splitter (PBS) into a coherent state
$|\alpha\rangle_{H}=||\alpha|e^{i\phi_{\alpha}}\rangle$ with
horizontal polarization and a coherent state
$|\beta\rangle_{V}=||\beta|e^{i\phi_{\beta}}\rangle$ with vertical
polarization. These coherent states are combined through a beam
splitter (BS1), producing two spatially separated beams, i.e, beam 1
and beam 2 at each output of the BS1. Beam 1 is sent to Bob and the
other beam 2 is kept by Alice. To create single photon quantum
channel between Alice and Bob, the beam 1 is attenuated to single
photon level, i.e., the mean photon number per pulse less than 1.
The single photon sent to Bob in the quantum channel is called
signal photon. The signal photon is inherited from the superposition
of $|\alpha\rangle_{H}+|\beta\rangle_{V}$ or two paths before the
BS1, where the relative phase
$(\phi_{\beta}-\phi_{\alpha})=-90^{\circ}$ is locked through beam 2
at Alice. Similarly, the beam 2 is further attenuated to single
photon level. The single photon kept in Alice is called ancilla
photon. The ancilla photon in Alice and the signal photon in Bob are
intrinsically correlated in the laser and phase-locked through the
phase-locking circuit at $(\phi_{\beta}-\phi_{\alpha})=-90^{\circ}$.
Then, the bi-partite correlation between these photons is obtained
by using wave plates in Alice and Bob as shown in the dotted boxes
in Fig.1. In the our previous work~\cite{sua11}, we have verified
four types of bi-partite correlation functions $\textrm{C1}=-
\textrm{cos} 2(\theta_{1}-\theta_{2})$, $\textrm{C2}=+\textrm{cos}
2(\theta_{1}+\theta_{2})$, $\textrm{C3}=-\textrm{cos}
2(\theta_{1}+\theta_{2})$, and $\textrm{C4}=+\textrm{cos}
2(\theta_{1}-\theta_{2})$ through the combination of wave plates as
shown in the inset of Fig.1. The half-wave plates(HWP3 and HWP5)
before the polarizing beam splitters (PBS1 and PBS2) in Alice and
Bob are used for projecting polarization angles $\theta_{1}$ and
$\theta_{2}$ so that maximum correlation, i.e.,
$\textrm{C}_{1,2,3,4}=\pm 1$ is obtained. In the following
discussion, let's assume that an ancilla photon and a signal photon
are available at the same time slots. We denote $'+'$ and $'-'$ for
these photons passed through and reflected from their PBSs. If the
single photon detector (SPD) $'+'$ or $'-'$ detects a photon, then
we encode the valid detection as bit $'1'$ or bit $'0'$,
respectively. For each correlation function, Alice and Bob can
control their photons to be in the $'+'$ or $'-'$ port of their PBSs
by using their HWPs. We need to choose the best settings for their
HWPs so that we could obtain maximum correlations of their photons
and also distribute the key effectively.

\begin{figure}
\includegraphics[scale=0.7]{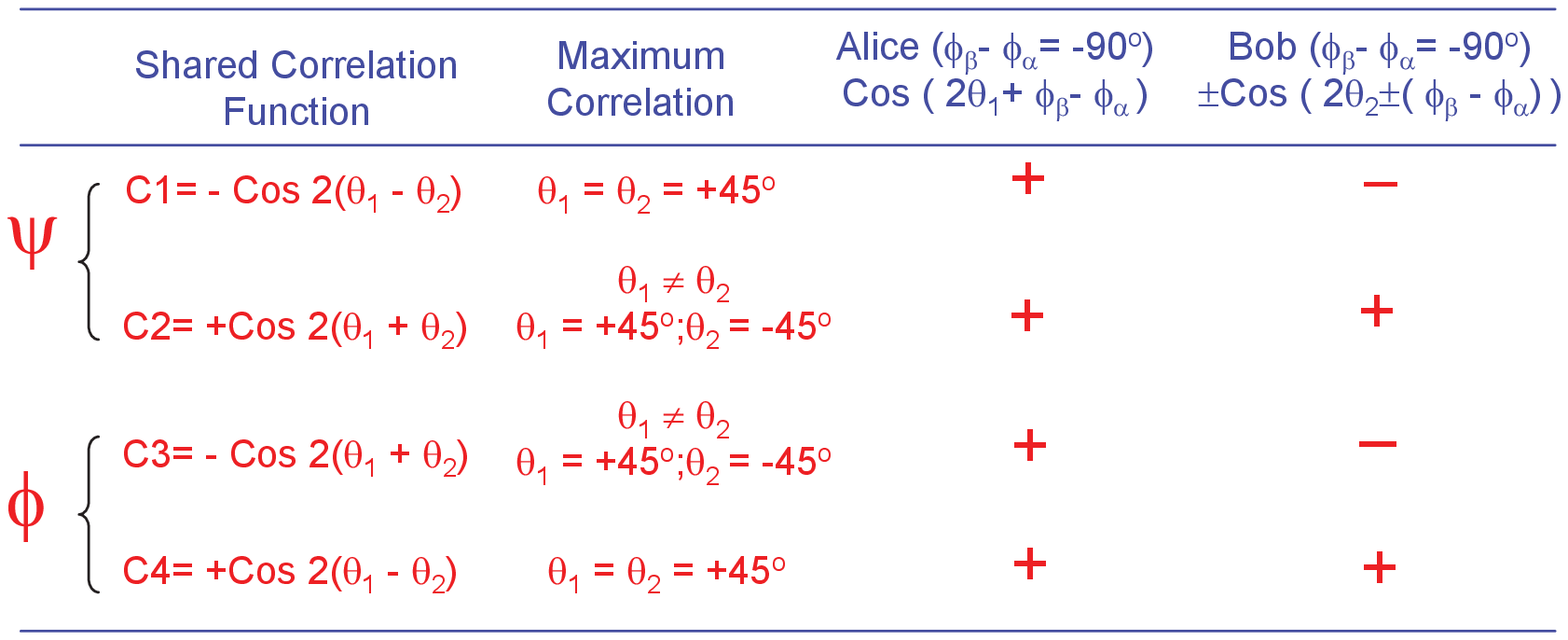}
\caption{The definition of the group ($\psi, \phi$), settings of
$\theta_{1}$ and $\theta_{2}$ for maximum correlation, the
interference signals at Alice and Bob for which detector will fire
for all four types of bi-partite correlations C1, C2, C3 and C4.}
\end{figure}

We illustrate each bi-partite correlation function and the optimum
settings of $\theta_{1}$ and $\theta_{2}$ in Fig.2. For the
correlation function $\textrm{C1}=-\textrm{cos}
2(\theta_{1}-\theta_{2})$, Alice and Bob receive their photons with
interference terms as given by
$+\textrm{Cos}(2\theta_{1}+(\phi_{\beta}-\phi_{\alpha}))$ and
$-\textrm{Cos}(2\theta_{2}+(\phi_{\beta}-\phi_{\alpha}))$,
respectively. By projecting the HWP5 in Alice at
$\theta_{1}=+45^{\circ}$ and also the HWP3 in Bob at
$\theta_{2}=+45^{\circ}$, we have C1=-1 indicating maximum
anti-correlation between Alice and Bob. This can be easily noticed
with the help of phase-locking
$(\phi_{\beta}-\phi_{\alpha})=-90^{\circ}$, the interference term in
Alice,
$+\textrm{Cos}(2(+45^{\circ})+(\phi_{\beta}-\phi_{\alpha}))\rightarrow
$ $'+'$ and interference term in Bob,
$-\textrm{Cos}(2(+45^{\circ})+(\phi_{\beta}-\phi_{\alpha}))\rightarrow
$ $'-'$. This implies that the ancilla photon in Alice passed
through the PBS2 and detected by the $'+'$ SPD3, and also the signal
photon in Bob is reflected from the PBS1 and detected by the $'-'$
SPD2. For the correlation function, $\textrm{C2}=+\textrm{cos}
2(\theta_{1}+\theta_{2})$, the maximum correlation C2=+1 is obtained
by projecting $\theta_{1}=+45^{\circ}$ and $\theta_{2}=-45^{\circ}$.
Alice has the same interference term, but Bob has the interference
term
$+\textrm{Cos}(2\theta_{2}-(\phi_{\beta}-\phi_{\alpha}))\rightarrow
+\textrm{Cos}(2(-45^{\circ})-(-90^{\circ}))\rightarrow$ $'+'$. As a
result, both the $'+'$ SPD3 in Alice and the $'+'$ SPD1 in Bob will
detect a photon. For the correlation function,
$\textrm{C3}=-\textrm{cos} 2(\theta_{1}+\theta_{2})$, we still
project $\theta_{1}=+45^{\circ}$ and $\theta_{2}=-45^{\circ}$ for
the maximum correlation C3=-1. Alice's $'+'$ SPD3 still see her
ancilla photon. While Bob has the interference term
$-\textrm{Cos}(2\theta_{2}-(\phi_{\beta}-\phi_{\alpha}))\rightarrow
-\textrm{Cos}(2(-45^{\circ})-(-90^{\circ}))\rightarrow$ $'-'$, and
so the $'-'$ SPD2 will see his signal photon. For the correlation
function, $\textrm{C4}=+\textrm{cos} 2(\theta_{1}-\theta_{2})$, the
$\theta_{1}=+45^{\circ}$ and the $\theta_{2}=+45^{\circ}$ are used
for the maximum correlation C4=1. Similarly, Alice will see her
ancilla photon in the '+' SPD3. Bob has the interference term
$+\textrm{Cos}(2\theta_{2}+(\phi_{\beta}-\phi_{\alpha}))\rightarrow
+\textrm{Cos}(2(45^{\circ})+(-90^{\circ}))\rightarrow '+'$. The '+'
SPD1 in Bob will detect a photon. For all four bi-partite
correlations, Alice will have her ancilla photon in the '+' SPD3.
Alice uses this valid detection for preparing the signal photon that
sent to Bob. We will divide the four bi-partite correlations into
two groups, C1,C2 $\rightarrow {\Psi}$ and C3,C4 $\rightarrow
{\Phi}$.

\begin{figure}
\includegraphics[scale=0.7]{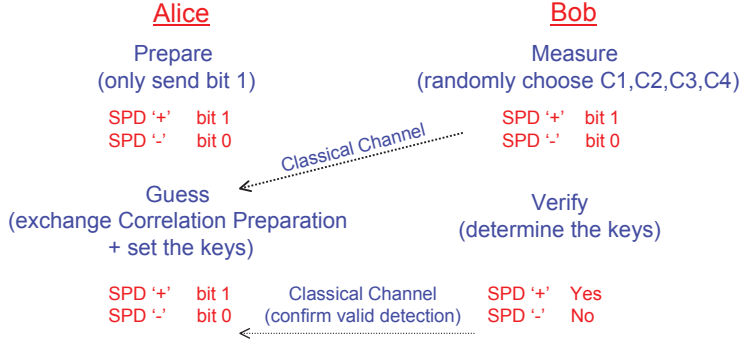}
\caption{Conceptual prepare-measure-guess-verify protocol for key distribution based on
single photon bi-partite correlation.}
\end{figure}

Now, let's discuss the protocol of key distribution between Alice
and Bob using the shared correlation function as shown in Fig.3. The
scheme can be divided into two parts; the first part is
Prepare-Measure (PM) part and the second part is Guess-Verify (GV).
In the PM part, Alice prepares an ancilla photon for herself and
also a signal photon to be sent to Bob. First, Alice has to verify
her ancilla photon is always bit $'1'$ or detected by the $'+'$ SPD3
as discussed before in Fig.2, so that the signal photon sent to Bob
is phase-locked, i.e, the relative phase between the horizontal and
vertical components of the signal photon is kept constant. From
here, we assume that Alice only prepares bit $'1'$ and sends the bit
information to Bob through the signal photon. Bob can randomly
choose one of the four C1, C2, C3, and C4 correlation functions by
means of randomly projecting the HWP1 and QWP2 as shown in the inset
of Fig.1. If Bob chooses C1, his $'-'$ SPD2 will 'click' as shown in
Fig.2 and then he encoded the detection as bit $'0'$. Similarly, if
Bob chooses C2 or C3 or C4, he will have bit $'1'$ or $'0'$ or
$'1'$, respectively. Bob can randomly generate the key by his choice
of correlation functions. However, the key is not shared with Alice.
Fig.2. shows the expected bit correlation for each correlation
between Alice and Bob in the PM part.

In the guess-verify (GV) part, Bob has to tell Alice about which
group $(\Psi, \Phi)$ of his choice by using a classical channel.
Since each group of $(\Psi, \Phi)$ has two choices of correlation
functions, Alice has to guess one of two correlations within the
group. No matter Alice's guess of Bob's choice of correlation is
right or wrong, Alice will use her guess of correlation to generate
the key bit. For example, Bob tells Alice that he used the group
${\Psi}$, Alice can chooses C1 or C2. If Alice choose C1 (C2), she
will generate bit $'1'$($'0'$) as her key bit. In order for Alice to
do that, she has to use the HWP1 and the QWP2 as shown in the dotted
box in Bob's setup in Fig.1. In the GV part, Alice is mimicking the
Bob's apparatus and generating the key bit based on her guess. She
send the signal photon to Bob. Bob will use the QWP4 at
$+45^{\circ}$ to replace the HWP1 and the QWP2 in his setup.
However, Bob must keep the setting of the HWP3 ($+45^{\circ},
-45^{\circ})$ for his choice of correlation function that he has
chosen in PM part. The sequence of the HWP3 angles will be used to
verify the key bit sent by Alice. The essence of this verify part is
Bob only (not Alice and Eve) knows his HWP3 angles. Bob can find out
the key generated by Alice's guess by detecting the signal photon in
the $'+'$ SPD1 ('Yes'/right guess) or the $'-'$ SPD2 ('No'/wrong
guess). To implement the GV part, Alice and Bob must keep their HWP
angles ($\theta_{1}=+45^{\circ}, +45^{\circ}, +45^{\circ},
+45^{\circ}; \theta_{2}=+45^{\circ}, -45^{\circ}, -45^{\circ},
+45^{\circ}$) for the correlation function C1, C2, C3 and C4,
respectively. Since Alice and Bob have swapped their correlation
preparation ($\textrm{HWP1+QWP2} \Leftrightarrow \textrm{QWP4}$) and
kept their projection angles (HWP3 and HWP5), Alice has to shift the
phase-locked mode to $\phi_{\beta}-\phi_{\alpha}=90^{\circ}$ for her
guess on the correlation functions C2 and C3. The reason is for the
correlation functions C2 and C3, Alice will have the interference
terms in the cosine function changed from
$+(\phi_{\beta}-\phi_{\alpha}))\rightarrow
-(\phi_{\beta}-\phi_{\alpha}))$.

\begin{figure}
\includegraphics[scale=0.8]{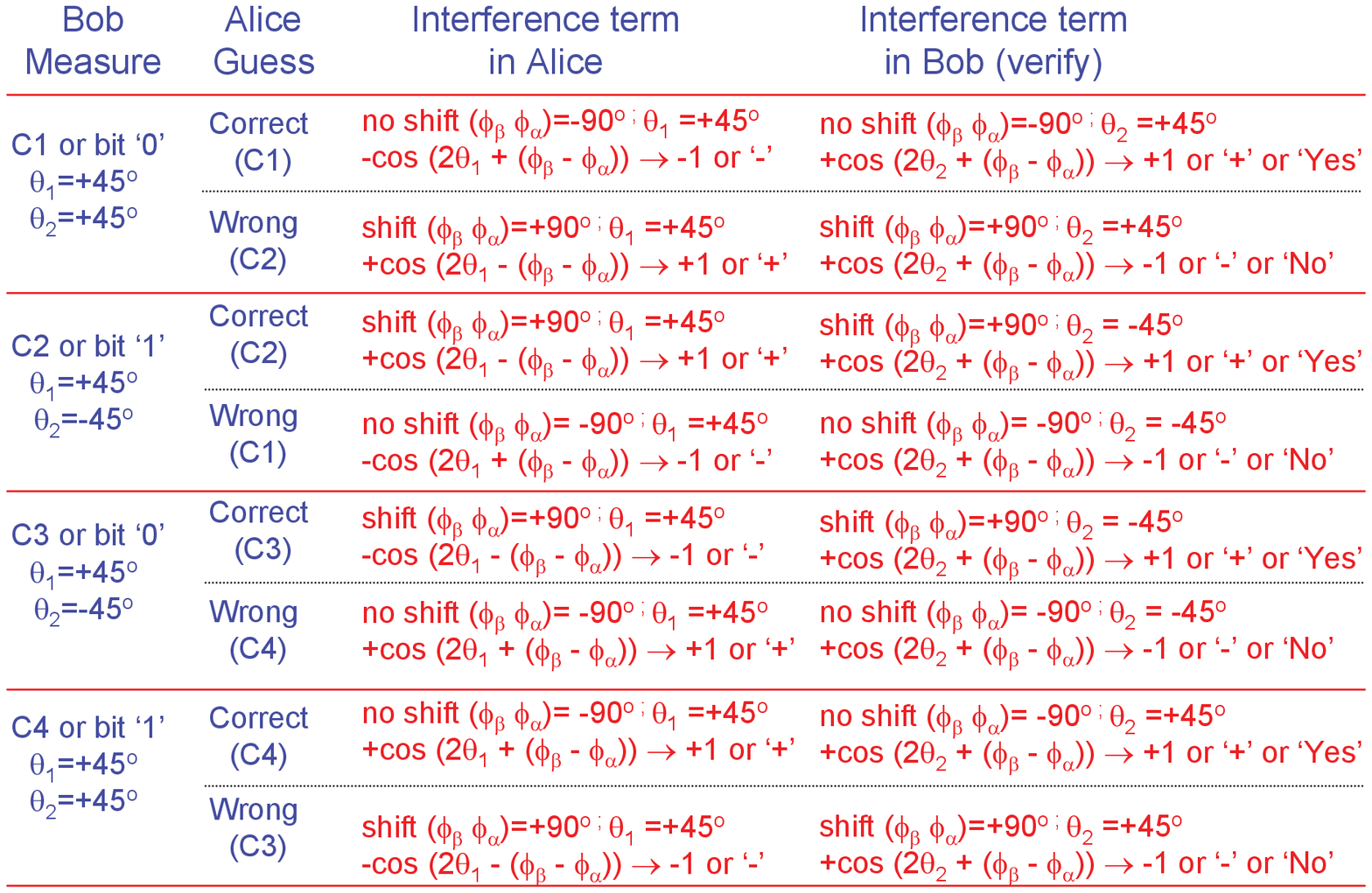}
\caption{The guess-verify part showing how Bob knows Alice's guess
is right or wrong for his choice of correlation in the
prepare-measure part. .}
\end{figure}

We illustrate the guess-verify part in more detail in Fig.4 about
how Bob knows Alice's guess is right or wrong. For the correlation
function C1, Alice and Bob keep their projection angles
$\theta_{1}=+45^{\circ}$ and $\theta_{2}=+45^{\circ}$ that they used
in the prepare-measure part. If Alice's guess on C1 is correct
through the group information $\Psi $ sent by Bob where Bob did
choose the C1 for his choice, the interference term in Alice
$-\textrm{cos}(2\theta_{1}+(\phi_{\beta}-\phi_{\alpha})))\rightarrow
-1$, i.e., bit $'0'$ or the $'-'$ SPD4 will detect the ancilla
photon. While the interference term in Bob
$+\textrm{cos}(2\theta_{2}+(\phi_{\beta}-\phi_{\alpha})))\rightarrow
+1$, i.e., the $'+'$ SPD1 will detect the signal photon which means
'Yes', so Bob knows that Alice has guessed the right correlation
function C1 and hence the bit $'0'$ for the key bit. Now, if Alice
guessed C2 instead of C1, so her guess is wrong. The interference
term in Alice
$+\textrm{cos}(2\theta_{1}-(\phi_{\beta}-\phi_{\alpha})))\rightarrow
+1$ or bit $'1'$. Note that Alice has to apply the phase shift
$\phi_{\beta}-\phi_{\alpha}=+90^{\circ}$ for her guess on C2 and C3
as discussed above. While the interference term in Bob
$+\textrm{cos}(2\theta_{2}+(\phi_{\beta}-\phi_{\alpha})))\rightarrow
-1$ or the $'-'$ SPD2 will 'click' which means 'No', so Bob knows
Alice has guessed the wrong correlation or bit. From here, Bob knows
the key bit set by Alice regardless Alice' guess is right or wrong.
Similarly, Bob knows Alice's guess for the other correlation
functions C2, C3 and C4 in the guess-verify part as illustrated in
Fig.4.

To illustrate the PMGV protocol more systematically, we will discuss
an example of the key distribution as shown in Fig.5. Step 1-4 is
for the PM part and Step 5-9 is for the GV part. Step 1: Alice sends
a phase-locked signal photon to Bob by making sure the ancilla
photon is detected in her $'+'$ SPD3 or bit $'1'$. Alice kept the
projection angle $\theta_{1}=+45^{\circ}$. Step 2: Bob measures the
signal photon based on his random choice of correlation function. He
chooses C3, C1, C4 and C2 and keeps the projection angle
$\theta_{2}$ for each correlation function. Step 3: He obtains bit
$'0'$, $'0'$, $'1'$, and $'1'$, respectively, according to Fig.2.
Step 4: Bob tells Alice through classical channel about the group of
his choice $(\Phi, \Psi)$, not revealing his choice of correlation
function. Step 5: Alice makes guess based on which group
information. Alice guesses C4, C1, C3, and C2. Step 6: Alice uses
the projection angle she kept in Step 1. She measures the ancilla
photon and obtains the bit $'1'$, $'0'$, $'0'$, and $'1'$ as her key
bit. Step 7: Bob measures the signal photon prepared by Alice's
guess by using the sequence of projection angle $\theta_{2}$ he kept
in Step 2. Bob knows whether Alice's guess is right ('Yes') or wrong
('No') as illustrated in Fig.4. Step 8a: Bob knows the key bits that
Alice set based on her guess even though Alice's guess was wrong.
Steps 8b and Step 9 are the alternative of Step 8a in the case Bob
did not receive the signal photon sent by Alice. Step 8b: Bob tells
Alice about his valid detection. The 'x' means no valid detection
and the '$\surd$' means valid detection. Step 9: Bob and Alice
shared the remain bits as their raw secret key.

\begin{figure}
\includegraphics[scale=0.7]{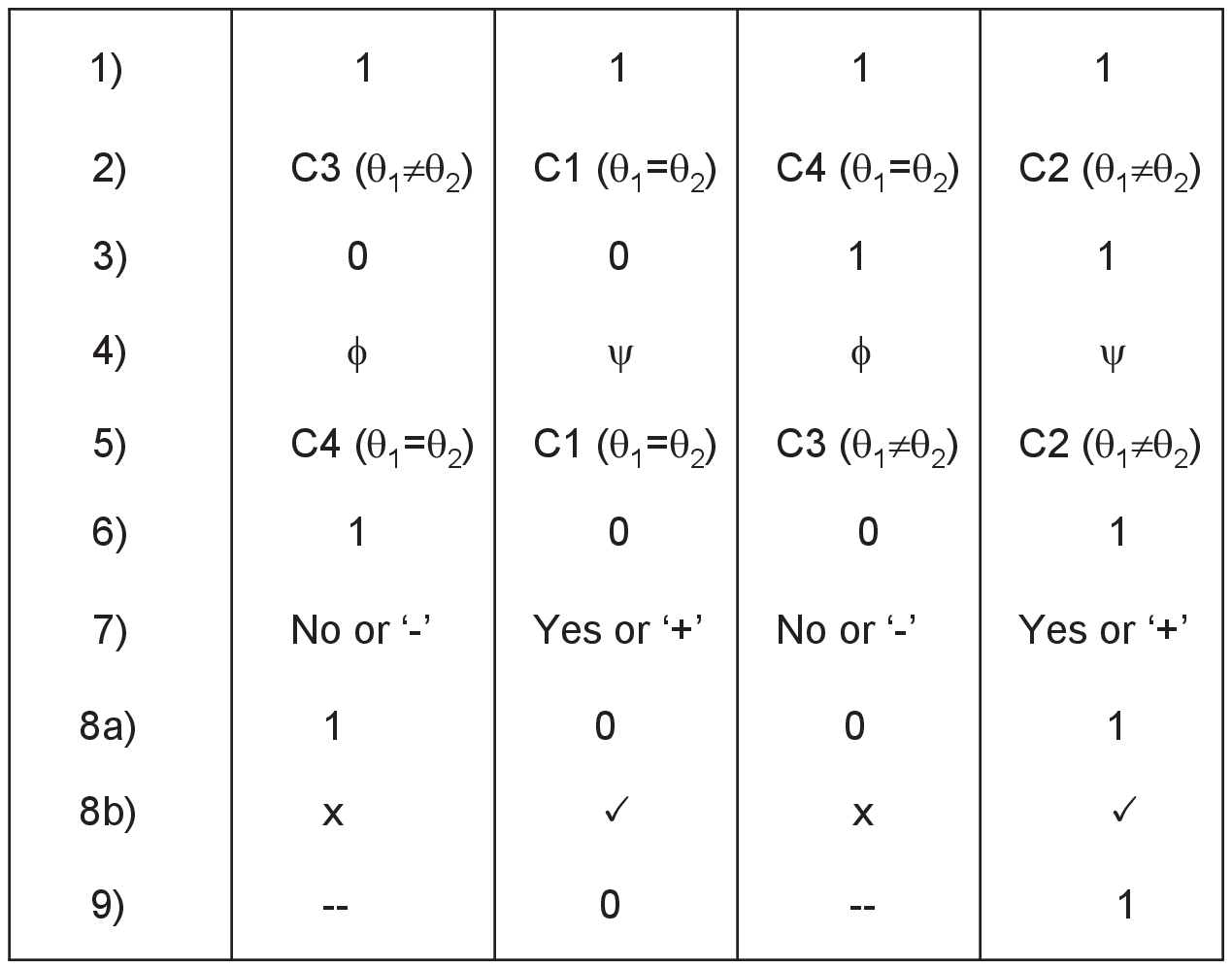}
\caption{The scheme for key distribution between Alice and Bob.}
\end{figure}

This protocol is based on the bi-partite correlation function
generated through the interference of the ancilla photon in Alice
and the interference of the signal photon in Bob. These
interferences are spatially separated but their phases are
intrinsically correlated in the laser. Since the signal photon is
prepared from a highly attenuated laser light, the protocol is still
vulnerable to PNS attack. The protocol is secure against PNS attack
if a correlated photon-pair is used as a photon source for replacing
the highly attenuated laser light. The correlated photon-pair is
much easier to generate and less sensitive to decoherence than an
entangled photon-pair. The high purity of correlated photon-pair at
the telecom wavelengths can be generated through a four wave mixing
process in a 300 m dispersion-shifted fiber (DSF). The coincidence
to accidental coincidence ratio (CAR) $>$ 100 has been achieved by
suppressing the spontaneous Raman scattering process in a DSF cooled
at the liquid nitrogen temperature 77K~\cite{kflee06}. In the four
wave mixing process, two pumps photon are annihilated to create
energy-time correlated signal-idler photon pair. The signal and
idler photons are separated from the pump photons by using a
cascaded wavelength division multiplexing (WDM). The signal photon
is projected to left circular polarization and sent to Bob.
Similarly, the idler photon is projected to right circular
polarization and sent to Alice. The right and left circular
polarizations of idler/signal photons are analog to the phase-locked
ancilla and signal photons when the highly attenuated laser light is
used. In the guess-verify part, the polarizations of idler/signal
photons are exchanged to right $\rightleftharpoons$ left for the
correlation functions C2 and C3.

Since our protocol requires coincidence detection of ancilla photon
and signal photon, the highly attenuated laser light can provide the
success rate of key distribution as given by $n_{a}n_{s}$, where
$n_{a}$ and $n_{s}$ are mean photon number per pulse for the ancilla
and signal photons. As for the use of a fiber based correlated
photon-pair, the protocol is complete secure against PNS attack. The
success rate for the key distribution is given by the production
rate of the photon-pair per pulse, $n_{pr}$. A cooled
dispersion-shifted fiber at liquid nitrogen temperature (77K) can
provide the purity of photon-pair with CAR $>$ 100 at photon-pair
production rate of 0.01 per pulse. For example, if we use the
commercial available $(\textrm{u}^{2}\textrm{t})$ fiber mode-locked
laser at repetition rate of 10 GHz~\cite{Liang07} and a high speed
low dark count super-conducting single photon detector, we will
obtain raw secret key bits about $0.01 \times 10^{9}\times
0.01\textrm{(total detection efficiency)}\sim 10^{5}$ key bits.
After applying private amplification and information reconciliation
protocols, we predict to obtain about 25000 secret key bits. If a
highly attenuated laser light is used to prepare the $n_{a}n_{s} =
0.01$ per pulse for both ancilla and signal, we can have the same
performance as discussed above. In ideal case, the best performance
of this protocol can be achieved by replacing highly attenuated
laser light or photon-pair source with two time-synchronized
deterministic single photon sources, where the $n_{a}n_{s} = n_{pr}=
1.0$ per pulse. It is worth to note that the photon-pair source and
two single-photon sources can increase distance of communication
between two parties by factor of two.

In conclusion, we have proposed a prepare-measure-guess-verify
(PMGV) protocol for key distribution using four types of single
photon bi-partite correlation functions between two parties. We show
that the PMGV protocol can be implemented with a highly attenuated
laser light source, which is often used as alternative single photon
source. Since the protocol requires coincidence detection of an
ancilla photon and a signal photon, any photon-pair source or two
single-photon sources can improve the success rate of key
distribution, the security against PNS attack and double the
communication distance in comparison to the use of highly attenuated
laser light.

\begin{acknowledgments}
K.F.L would like to acknowledge the financial support as visiting
professorship from University of Malaya. H.B.A gratefully
acknowledges the support from the University of Malaya High Impact
Research Grant (UM.C/HIR/MOHE/SC/01) on this work.

\end{acknowledgments}


\newcommand{\noopsort}[1]{} \newcommand{\printfirst}[2]{#1}
  \newcommand{\singleletter}[1]{#1} \newcommand{\switchargs}[2]{#2#1}

\end{document}